\documentstyle[11pt,hrd_pasp,epsfig,twoside]{article}
\begin{document}

\title{The Color-Magnitude Diagram of composite stellar populations}
\author{Laura Greggio}
\affil{Osservatorio Astronomico, Via Ranzani 1, 40127 Bologna, Italy}

\begin{abstract}

The theoretical relation between the number of post main sequence
stars from stellar populations and their total mass is 
investigated. This is used to derive some 
basic relations between the stellar number counts of stellar populations
with an age spread and the star formation history which produced them.
The main purpose of this investigation is to offer a guideline for the 
construction of simulations aimed at reproducing color-magnitude diagrams.

\end{abstract}

\section{Introduction}

Theoretical simulations of Color-Magnitude Diagrams (CMD) have proven
very successful to investigate the Star Formation History (SFH) of nearby 
galaxies (see the contributions by Aparicio and by Tosi et al. in this volume).
In essence, the method works because, on the basis stellar evolution theory,
we relate star counts in selected regions of the CMD to the 
mass of the stellar population which generated them. Since different regions 
of the CMD are populated by stars of different ages, the stellar counts
speak for the mass that went into stars as a function of time.
However, there are regions in the CMD which, sampling stars from a 
large age range, only yield 
integrated information. In this paper I will anticipate some 
results of an investigation of the potentials and the limitations of the 
stellar evolution theory for reconstructing the SFH. 
I will consider stellar populations with an age spread, but a fixed 
metallicity. I will also just focus on the number counts of 
stars in Post Main Sequence (PMS)
evolutionary phases. Often theses star occupy the brightest portions of the
observational CMDs. In addition, for this kind of stars a straightforward 
relation applies between the counts and the mass of the parent population.

\section{The Number-Mass connection}

For an Simple Stellar Population (SSP), i.e. a collection of single stars
all with the same age ($\tau$) and metallicity ($Z$), 
the number of stars in the $j$-th
PMS evolutionary phase is well approximated by (Renzini 1981):

\begin{equation}
\Delta N_{j}^{SSP} = \phi(M_{TO})\, |\dot{M}_{TO}|\, \Delta t_{j}
\end{equation}

\noindent
where $M_{TO}$ to is the turn off mass, $\dot{M}_{TO}$ is its time 
derivative, $\Delta t_{j}$ is 
the lifetime of the PMS phase considered, and $\phi(M_{TO})$ is the Initial 
Mass Function (IMF) by number evaluated at the turn-off mass.

I characterize the SSP for its total mass in the range 
0.6$-$120 $M_\odot$ ($M_{>0.6}^{SSP}$):

\begin{equation}
M_{>0.6}^{SSP}= \int_{0.6}^{120} \phi(M) \,dM = 
A\,\int_{0.6}^{120} M^{1-\alpha}\,dM = A\, / \, f_\alpha
\end{equation}

\noindent
having assumed a single slope power law for the IMF in the range 0.6$-$120 
$M_\odot$. In this notation, Salpeter IMF corresponds to $\alpha$=2.35

The scaling of the number of PMS stars with the mass of the parent population
then becomes:

\begin{equation}
\Delta N_{j}^{SSP} = M_{>0.6}^{SSP} \, \zeta_{\alpha} \, \Delta t_{j} = 
M_{>0.6}^{SSP} \times \Delta n_{j}^{SSP}
\label{eq:dNjs}
\end{equation}

\noindent
where \(\zeta_{\alpha}\,=f_{\alpha}\,M_{TO}^{-\alpha}\,|\dot{M}_{TO}|\) 
is the \textit{specific evolutionary flux}, i.e. the number of stars 
leaving the MS per unit time per unit mass of the parent population.
The quantity $\Delta n_{j}^{SSP}$ is the 
\textit{specific production of PMS stars} 
in the $j$-th evolutionary phase of the SSP with age $\tau$ and metallicity 
$Z$, such that its turn off mass is $M_{TO}$. All factors in 
equation ~\ref{eq:dNjs}  depend on $\tau$ and $Z$;
the dependence on the IMF is included in $\zeta_{\alpha}$.

\begin{figure}
\plotone{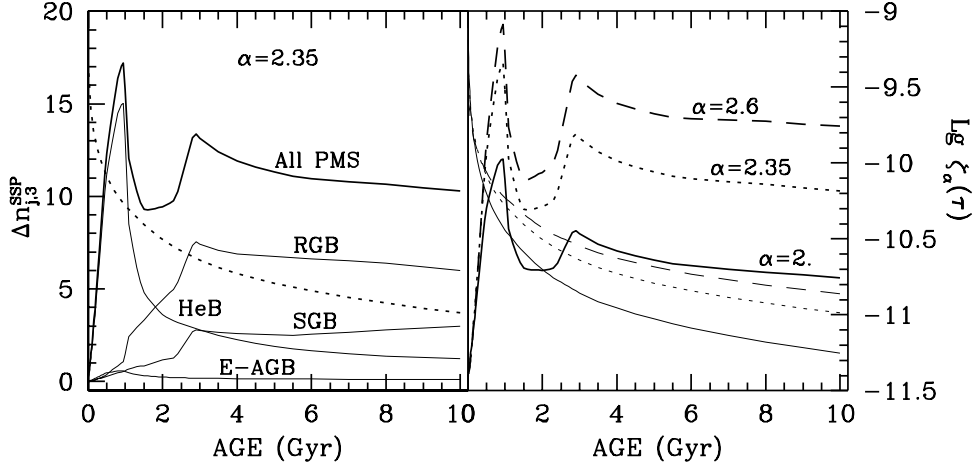}
\caption{Left panel: the production of PMS stars of SSPs with 
$M_{>0.6}^{SSP}$=1000M$_\odot$ for $\alpha$=2.35. The thick line refers to the
total PMS; the thin lines to the individual phases: 
sub-giant branch (SGB), red giant branch (RGB), central Helium burning (HeB),
double shell burning up to the first thermal pulse (Early AGB, E-AGB). The
dotted line shows the specific evolutionary flux (right scale).
Right panel: total PMS production (thick lines) and specific evolutionary 
flux (thin lines) for three values of $\alpha$. The linetype encoding is    
labelled.}
\label{f:nmass}
\end{figure}

The behaviour of the specific production of PMS stars as a function of
the SSP age is shown in Fig. \ref{f:nmass} for the set of tracks by Fagotto et
al. (1994) with Z=0.004. The trend with age results from two competing 
effects: the temporal decrease of the specific evolutionary flux as the
population ages (dotted line in the left panel), and the increase of the 
duration of specific PMS stages.
   
For young SSPs (up to $\approx$ 1Gyr) the specific production of PMS objects
increases: the lengthening of the lifetimes prevails, and it's by far
driven by the helium burning lifetime.

There is a sharp maximum in $\Delta n_{j}^{SSP}$ at $\approx$ 1 Gyr: 
this is due to the fact that SSPs around this age have the longest helium 
burning lifetime (e.g. Sweigart, Greggio, \& Renzini 1990; Girardi \&
Salaris 2001). 

A second maximum appears at $\approx$ 3 Gyr, caused by the
lengthening of the RGB lifetime. After this epoch the specific production of 
PMS stars mildly decreases with increasing age, as the RGB lifetime levels 
off. 

The right panel of Fig. \ref{f:nmass} illustrates the impact of the IMF
slope on $\Delta n_{j}^{SSP}$. Older than a few
hundred Myr, steeper slopes correspond to 
a larger number of PMS stars at fixed $M_{>0.6}^{SSP}$, reflecting the 
larger stellar evolutionary flux.

\begin{figure}
\plotone{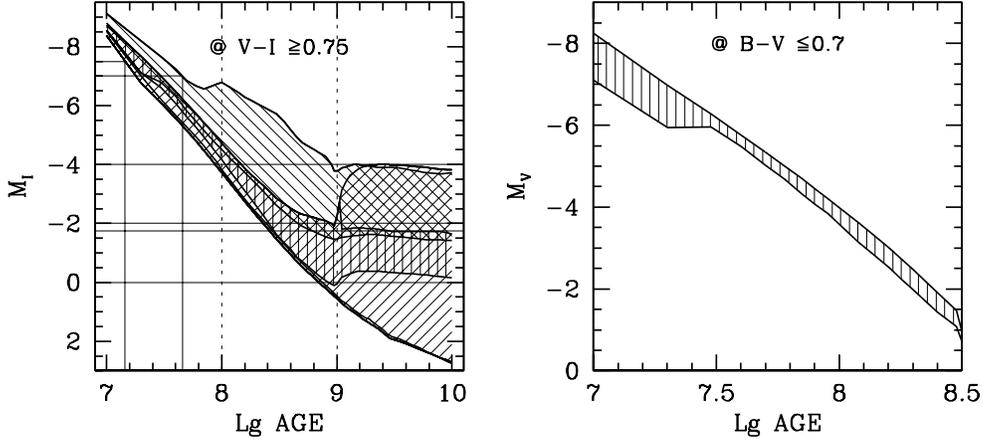}
\caption{Magnitude range covered by red (left panel) and blue (right panel)
PMS stars of SSPs with ages between 10 Myr and 10 Gyr. The color selection
is labelled in each panel. Vertical shading
refers to HeB; slanted shading to RGB (angle=+45\deg) and E-AGB 
(angle=-45\deg) stars.}
\label{f:taumag}
\end{figure}

\smallskip
I now pass to consider the location in the CMD where these PMS stars are to 
be counted. Fig. \ref{f:taumag} shows the magnitude range covered by 
PMS stars of SSPs with ages from 10 Myr to 10 Gyr. The
Fagotto Z=0.004 tracks in combination with the Bessel, Castelli and Pletz 
(1998) model atmospheres with \([M/H]=-0.5\) have been used to produce 
this plot. Different shading refer to different PMS phases (see caption). 

There is a well defined $M_I - \tau$ relation for helium 
burners up to $\approx$ 1 Gyr, that is the age at which the Red Giant Branch
develops to full extension (RGB phase transition). Thereafter the He burning 
is spent at virtually the same luminosity, indipendent on age. 

In the intermediate age regime, the luminosity at the 1st AGB thermal pulse
can be used to age-date the SSPs. This is at the basis of the
AGB-tip age dating, although the location of the tip itself is
uncertain due to our poor knowledge of the mass loss processes
during the AGB evolution.

Beyond 1--2 Gyrs there's no bright feature that betrayes the
age of the SSP. To age-date the SSP we need to sample as deep as the 
RGB base. A notable exception is the presence of blue horizontal branch
stars, but the corresponding age-dating is uncertain because the color 
distribution of these stars is sensitive to various poorly understood effects
(see e.g. the second parameter problem (Renzini \& Fusi Pecci 1988)).

\section {Composite Stellar Populations}

Having introduced the basic ingredients, I now pass to examine the connection
between stellar counts in magnitude bins on the CMD of stellar populations
with an age range, and their mass. For such Composite Stellar Population
(CSP) eq. (\ref{eq:dNjs}) becomes:

\begin{equation}
\Delta N{_j}^{CSP} = \int_{\tau_n}^{\tau_x}\dot{M}(\tau)\,\Delta n_{j}^{SSP}
\, d\tau
\label{eq:deltaN}
\end{equation}

\noindent
where the subscript $j$ now refers to the magnitude bin, $\dot{M}(\tau)$ is
the star formation rate (SFR) at the age $\tau$, and the integration is 
performed over the age range contributing to the counts in the $j$-th bin.

The brightest regions of the CMD sample a relatively small range of ages.
In this regime:

\begin{equation}
\Delta N_{j} \simeq <\dot{M}>_{\tau_j}\,<\Delta n>_{\tau_j}\,\Delta \tau_j
\end{equation}
 
\noindent 
where the SFR and the specific productions are evaluated at the typical 
age $\tau_j$ sampled by the $j$-th magnitude bin, and $\Delta \tau_{j}$ 
is the intercepted age range. As long as $\Delta \tau_{j}$ is small
(i.e. the position of the stars on the CMD varies steeply with their
age) the luminosity function (LF) gives a detailed description of the
star formation rate. Inspection of the left panel of Fig. \ref{f:taumag} 
shows that this
is the case only for the brightest portion of the red LF. Magnitude bins
fainter than $M_I \sim -6$ already collect E-AGB stars from a large age range.
This problem is not present in the blue part of the CMD (right panel of 
Fig. \ref{f:taumag}), where $\Delta \tau_{j}$ remains
small up to a few hundred Myrs. Thus counts of blue helium burners
give a fair indication of the SFR over this age range (Dohm-Palmer et al.
1997).  

\begin{figure}
\plotone{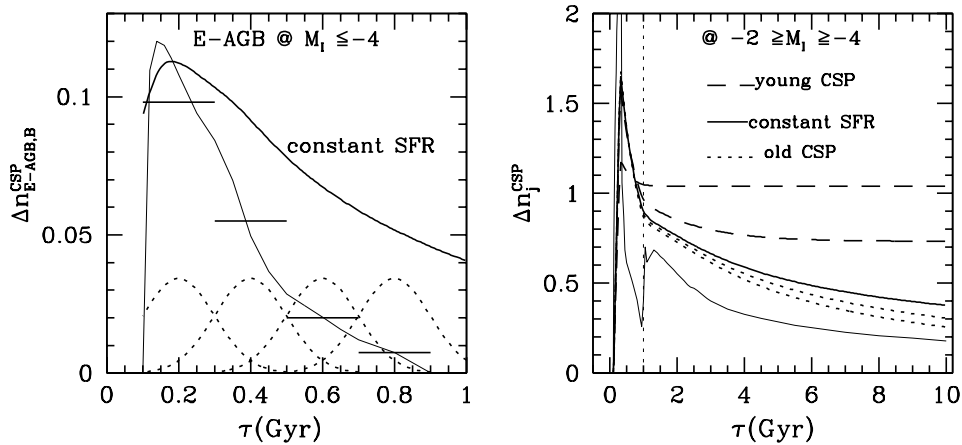}
\caption{Production of stars from SSPs (thin lines)
and CSPs (thick lines) with $M_{>0.6}^{SSP}$=1000M$_\odot$ and $\alpha$=2.35
The left panel refers to E-AGB stars brighter than the RGB Tip, which are 
produced at ages between 0.1 and 0.9 Gyr in the adopted tracks. The right
panel shows the production of stars with $V-I \ge 0.75$ and 
$-2 \ge M_I \ge -4$. See text for more details.}
\label{f:nspec}
\end{figure}

\medskip
For older stellar populations, the bright portion of the CMD yields basically
integrated information, as can be appreciated from Fig. \ref{f:taumag}. 
I turn then to consider the specific production of objects in a region of the
CMD from a CSP:

\begin{equation}
\Delta n_j^{CSP} = \frac{\int_{\tau_n}^{\tau_x}\dot{M}(\tau)\,
\Delta n_{j}^{SSP}\, d\tau}{\int_{\tau_n}^{\tau_x}\dot{M}(\tau)\,d\tau} = 
<\Delta n_{j}^{SSP}>_{\dot{M}_\tau}
\label{eq:deltan}
\end{equation}

The specific production of $j$-type stars of a composite stellar population 
with a range of ages is the weighted mean of the specific productions of
the SSPs, with the weights given by the SFR in the relevant age range.
The star counts in a region of the CMD 
divided by $\Delta n_j^{CSP}$ yields the mass that went into stars in the
age range corresponding to the selected region of the CMD. It is thus 
important to see how this quantity varies under different assumptions for
the SFH. In the following I will consider the stellar counts in the red 
portion of the CMD.

The left panel of Fig. \ref{f:nspec} illustrates the specific production 
of E-AGB stars brighter than
the RGB tip (in the I band) for a variety of SF histories in the age range 
from 0.1 to 0.9 Gyr.
The thick horizontal lines show the level of $\Delta n_j^{CSP}$
for bursts of SF, modelled as gaussian distributions centered on progressively
older ages, all with a 0.1 Gyr age dispersion (dotted lines). 
The thick continuous line, instead, shows $\Delta n_j^{CSP}$
as a function of the maximum age of the CSP, for
a family of SF episodes all occurred at a constant rate up to
$\tau_n = 0.1$ Gyr.

$\Delta n_j^{SSP}$ decreases rapidly with age: 
the older the stellar population, the more mass we have to put into stars 
to account for a given observed number of bright E-AGB stars. 

The ratio between the 
mass of the CSP and the star counts depends on the age and the age limits
of the SF episode considered. On the other hand the luminosity and colour 
of the AGB stars can be used to rank their age distribution, 
so that the uncertainty in the derived CSP mass can be estimated within a 
factor of a few.

The right panel of Fig. \ref{f:nspec} shows the specific production of 
stars in 
the bin $-2 \ge M_I \ge-4$, which collects young helium burners, E-AGB
intermediate age and old (AGB + RGB) bright stars (see Fig. \ref{f:taumag}).
In this magnitude bin, $\Delta n_j^{SSP}$ is very large for the young 
Helium burners, 
temporarily drops at intermediate ages (where the SSP only yield AGB stars),
and has a second peak at the RGB phase transition. For ages
older than $\approx$ 1Gyr, $\Delta n_j^{SSP}$
is only mildly varying  over the whole range of ages considered. 
It follows that, in $-2 \ge M_I \ge-4$, the specific production of stars 
from CSP with 
$\tau_{n} \ge $ 1 Gyr only marginally depends on the age range. 

The impact of the presence of young stars ($\tau_n \le 1$Gyr) in this 
magnitude bin is illustrated by the thick lines for a variety of SFHs. 
Specifically,
the dashed lines correspond to an age distribution $\propto e^{-\tau/\tau_0}$;
the solid line to a constant SFR; the dotted line to an age distribution
$\propto e^{\tau/\tau_0}$. The ratio of mass in stars younger than 1 Gyr to
the mass in stars with ages between 1 and 10 Gyr is 5 (upper dashed), 1
(lower dashed), 0.1 (solid), 0.05 (upper dotted) and 0.02 (lower dotted).
It can be seen that, if the CSP maximum age is
older that the RGB phase transition (which gives well defined signature on the
CMD, i.e. the RGB tip), the specific production is not very sensitive to
the SF history and age limits. Again, the uncertainty of the conversion
of stellar counts to total mass of the parent population can be estimated
around a factor of a few.
Similar considerations hold for a magnitude range which samples the
helium burners (e.g. $0 \ge M_{I} \ge -1.75$ in Fig. \ref{f:taumag}). 
 
\medskip

To summarize: 

\begin{itemize}
\item The bright portion of the LF of PMS stars allows to recover the 
SFH with a fair degree of detail, up to ages of a few 10$^8$ years.
\item For older ages, the specific production of stars in the various
regions of the CMD depends on the age limits and on the SFR law of the 
CSP. It is however possible to estimate the total mass of the CSP within 
a factor of a few, for an assumed IMF slope. 
\end{itemize}

On the average, for Salpeter IMF one gets approximately 
1 E-AGB stars brighter than the RGB Tip every 20000 $M_\odot$,
1 (red) star in $-2 \ge M_{I} \ge -4$ every 2000 $M_\odot$
and 1 (red) star star in $0 \ge M_{I} \ge -1.75$ 
every 200 $M_\odot$ of the parent population (in stars more massive than 
0.6 $M_\odot$). 

The best way to decode the stars' distribution on the 
CMD in terms of SFH is by computing synthetic CMDs, which account for the 
stellar evolution, number statistics and observational uncertainties.
In addition, the simulations allow to benefit of the information from the
MS phase, and to cross check the star counts in different CMD regions which 
sample the same ages. Nevertheless, the Number-Mass relation illustrated
here offers a quick-look tool to picture the SF history from an observational 
CMD, as well as theoretical support for the results we achieve through
the construction of synthetic CMDs.

\end{document}